\documentclass[aps,prl,twocolumn,amsmath,amssymb,floatfix,superscriptaddress]{revtex4-1}
\usepackage{tabularx}
\usepackage{bm}
\usepackage{epsfig,psfrag,subfigure}
\usepackage{graphicx}
\usepackage{color}
\usepackage{amsfonts}
\usepackage{exscale}
\usepackage{amsbsy}
\usepackage{multirow}

\usepackage{times}
\usepackage{float}
\usepackage{latexsym,amsmath,amssymb,bm,euscript}
\usepackage{epstopdf}
\usepackage[colorlinks=true,linkcolor=blue,citecolor=blue]{hyperref}
\usepackage{soul}
\usepackage[normalem]{ulem}
\usepackage{mathrsfs}
\usepackage{amsmath}
\usepackage{lettrine}
\usepackage{bbding}
\usepackage{xspace}
\usepackage{textcomp}
\usepackage{textcase}
\usepackage{setspace}

\usepackage[dvipsnames]{xcolor}
\def\be{\begin{equation}}       \def\ee{\end{equation}}
\def\bea{\begin{eqnarray}}      \def\eea{\end{eqnarray}}

\newcommand{\PreserveBackslash}[1]{\let\temp=\\#1\let\\=\temp}
\newcolumntype{C}[1]{>{\PreserveBackslash\centering}p{#1}}

\begin{document}
\title{Enhanced superconductivity by near-neighbor attraction in the doped Hubbard model}

\author{Cheng Peng}
\affiliation{Stanford Institute for Materials and Energy Sciences, SLAC National\\Accelerator Laboratory, Menlo Park, California 94025, USA}

\author{Yao Wang}
\affiliation{Department of Physics and Astronomy, Clemson University, Clemson, SC 29631, USA}

\author{Jiajia Wen}
\affiliation{Stanford Institute for Materials and Energy Sciences, SLAC National\\Accelerator Laboratory, Menlo Park, California 94025, USA}

\author{Young S. Lee}
\affiliation{Stanford Institute for Materials and Energy Sciences, SLAC National\\Accelerator Laboratory, Menlo Park, California 94025, USA}
 \affiliation{Department of Applied Physics, Stanford University, Stanford, California 94305, USA}

\author{Thomas P. Devereaux}
\affiliation{Stanford Institute for Materials and Energy Sciences, SLAC National\\Accelerator Laboratory, Menlo Park, California 94025, USA}
\affiliation{Department of Materials Science and Engineering, Stanford University, Stanford, California 94305, USA}

\author{Hong-Chen Jiang}
\email{hcjiang@stanford.edu}
\affiliation{Stanford Institute for Materials and Energy Sciences, SLAC National\\Accelerator Laboratory, Menlo Park, California 94025, USA}

\begin{abstract}
Recent experiment has unveiled an anomalously strong electron-electron attraction in one-dimensional copper-oxide chain Ba$_{2-x}$Sr$_x$CuO$_{3+\delta}$. While the effect of the near-neighbor electron attraction $V$ in the one-dimensional extended Hubbard chain has been examined recently, its effect in the Hubbard model beyond the one-dimensional chain remains unclear. Here, we report density-matrix renormalization group studies of the extended Hubbard model on long four-leg cylinders on the square lattice. We find that the near-neighbor electron attraction $V$ can notably enhance the long-distance superconducting correlations while simultaneously suppress the charge-density-wave correlations. Specifically, for a modestly strong electron attraction, the superconducting correlations become dominant over the CDW correlations with a Luttinger exponent $K_{sc}\sim 1$ and strong divergent superconducting susceptibility. Our results provide a promising way to realize long-range superconductivity in the doped Hubbard model in two dimensions. The relevance of our numerical results to cuprate materials is also discussed.
\end{abstract}

\maketitle

\date{February 2022}

\maketitle

The origin of unconventional superconductivity is one of the greatest mysteries since the discovery of high-$T_c$ cuprates\,\cite{bednorz1986possible}. Contrary to the conventional BCS superconductors, it is widely believed that the strong electronic Coulomb repulsions in the $3d$ orbitals play the dominant role in the $d$-wave pairing mechanism in cuprates. Along this line, spin fluctuations generated by the doped antiferromagnetic state may provide the pairing glue\,\cite{scalapino1986d,gros1987superconducting,kotliar1988superexchange,tsuei2000pairing}. Based on the minimal model describing the correlations effect -- the single-band Hubbard model\,\cite{zhang1988effective,natphyhubbard2013,Keimer2015nat,Srinivas2022Hubbard,Gull2021Hubbard}, this pairing mechanism has been proposed based on perturbation theory and instabilities on small clusters\,\cite{anderson1987resonating,dagotto1994correlated}. However, the ultimate verification requires the exact proof of long-range orders $d$-wave superconductivity in the thermodynamic limit.

To address these questions, advanced numerical simulations have been applied to the Hubbard model and its low-energy analog -- the $t$-$J$ model, in the past few years\cite{Gull2021Hubbard,Srinivas2022Hubbard}. Many unusual phases in cuprates, such as a striped phase\,\cite{zheng2017stripe, huang2017numerical,huang2017science,huang2018stripe,ponsioen2019period} and a strange metal phase\,\cite{kokalj2017bad, huang2019strange, cha2020slope}, have been identified recently by unbiased and exact methods on clusters such as density matrix renormalization group (DMRG) and determinantal quantum Monte Carlo (DQMC). However, the search for $d$-wave superconducting (SC) phase is not been quite as successful.\cite{Mingpu2020prx} Quasi-long-range SC order has been found in the Hubbard and $t$-$J$ models on four-leg square lattice cylinders,\cite{Jiang2019Hub,Jiang2020prr,Chung2020,Dodaro2017,Jiang2018tJ,jiang2021ground,Jiang2021prl,Gong2021,Jiang2020prb} that may be tuned by the band structure and in the striped Hubbard model\cite{Jiang2022pnas}. However, when a similar study was extended to the wider six and eight-leg $t$-$J$ cylinders on the square lattice, which is closer to two dimensions, superconductivity is found to disappear in the hole-doped side\,\cite{jiang2021ground,Jiang2021prl,Gong2021}. Therefore, the original Hubbard and $t$-$J$ models themselves might not be sufficient to resolve the high-$T_c$ puzzle.

Meanwhile, experimental efforts have been devoted to the search of new insights. In a very recent photoemission experiment, a strong near-neighbor electron attractive interaction was identified in 1D cuprate chains, which may be mediated by phonons\,\cite{chen2021anomalously,wang2021phonon}. Such an interaction is likely to be a missing ingredient also in high-$T_c$ cuprates. Intuitively, the extended Hubbard model (EHM) with on-site repulsion and near-neighbor electron attraction may favor nonlocal Cooper pairs\,\cite{jiang2022enhancing,huang2021superconductivity}. Moreover, a recent DMRG study has identified dominant $p$-wave SC correlations in the pairing channel in the one-dimensional (1D) EHM\,\cite{Qu2021arxiv}. These recent experimental and theoretical discoveries motivate the investigation of $d$-wave superconductivity with the presence of a near-neighbor attractive electron interaction.

\textit{Principal results --}
Previous DMRG studies \cite{Jiang2019Hub,Jiang2020prr,Chung2020} have shown that the ground state of the lightly doped Hubbard model on four-leg square cylinders with next-nearest-neighbor (NNN) electron hopping $t'$ is consistent with that of a Luther-Emergy liquid (LE)\cite{Luther1974} 
 which is characterized by quasi-long-range SC and charge-density-wave (CDW) correlations but exponentially decaying spin-spin and single-particle correlations. However, while both the SC and CDW correlations are quasi-long-ranged, the former dominates over the latter in all cases, which suggests that CDW order may be realized in the two-dimensional (2D) limit. In this paper, we show that the presence of a finite nearest-neighbor (NN) electron attraction $V$ can notably enhance the SC correlations, while suppressing the CDW correlations simultaneously. This demonstrates the mutual competition relation between the SC and CDW orders in the Hubbard model. More importantly, we find that the SC correlations become dominant over the CDW correlations when the electron attraction is modestly strong. This suggests that the SC order, instead of the CDW order, may be realized in the 2D limit when NN attractions are present. Our results provide a promising pathway to potentially realize long-range superconductivity in the Hubbard model.

\begin{figure}[tb]
\centering
    \includegraphics[width=1\linewidth]{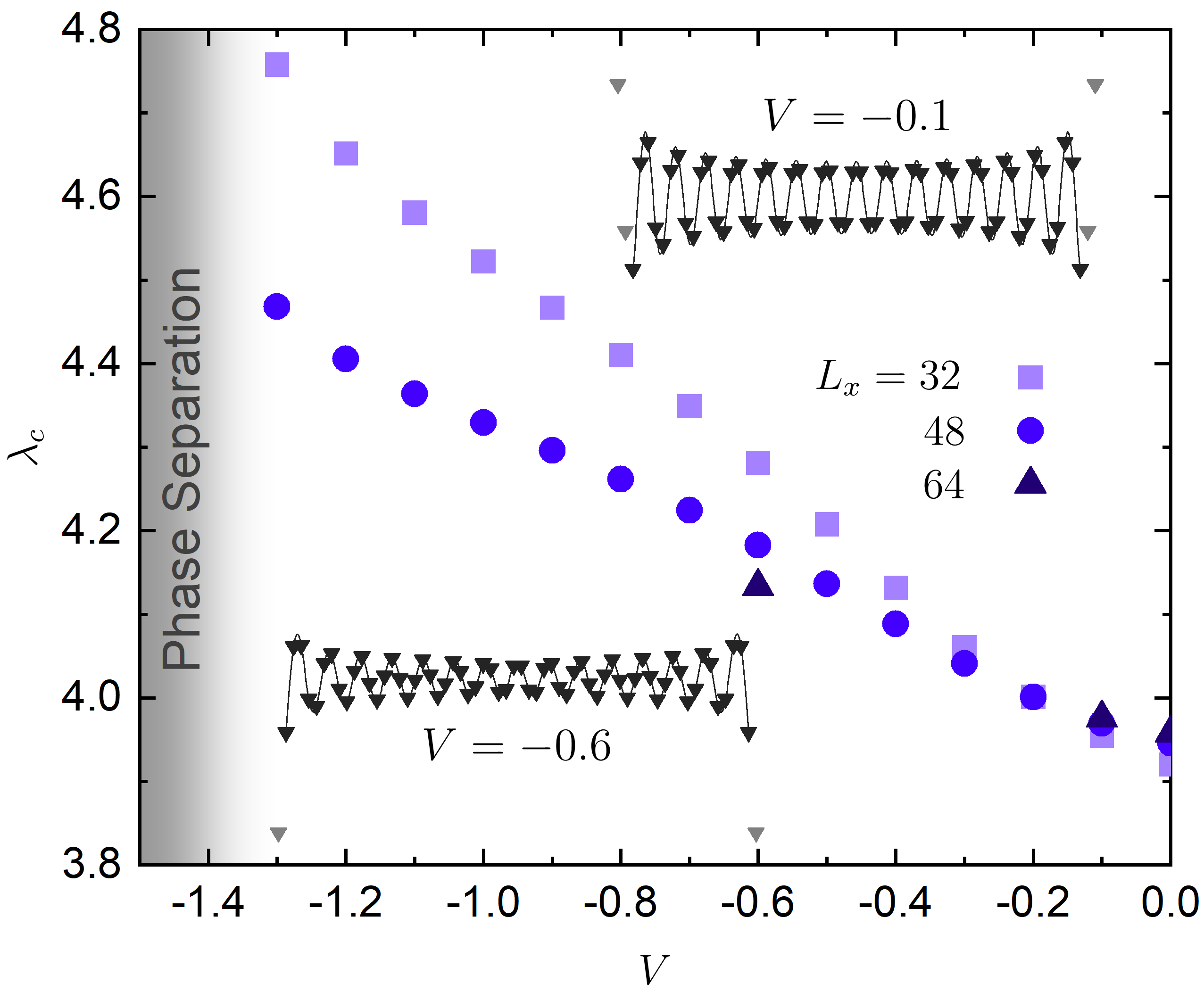}
\caption{(Color online) The CDW wavelength $\lambda_c$ as a function of $V$. The length of the cylinder is $L_x=32$, $48$, and $64$, and the hole doping concentration is $1/8$. Inset: charge density profiles for $V=-0.1$ and $-0.6$, respectively. The fitting curves are obtained using Eq.\ref{Eq:FriedelOscillation}. Note that a few data points close to the boundaries (in grey) are removed in the fittings to minimize boundary effects. The gray shaded area denotes phase separation. 
} \label{Figs:Q-V}
\end{figure}

\textit{Model and Method --} %
We employ the DMRG method\cite{White1992} to study the ground state properties of the single-band extended Hubbard model on the square lattice, defined by the Hamiltonian%
\begin{eqnarray} \label{Eq:Ham}
  H= &-& \sum_{ij,\sigma}t_{ij}(\hat{c}^{\dagger}_{i\sigma}\hat{c}_{j\sigma}+h.c.) + U\sum_{i}\hat{n}_{i\uparrow}\hat{n}_{i\downarrow} \nonumber\\
  &+& V\sum_{\langle ij \rangle}\hat{n}_{i} \hat{n}_{j}.
\end{eqnarray}
Here, $\hat{c}^{\dagger}_{i\sigma}$ ($\hat{c}_{i\sigma}$) is the electron creation (annihilation) operator with spin-$\sigma$ ($\sigma=\uparrow, \downarrow$) on site $i=(x_i,y_i)$, $\hat{n}_{i\sigma}=\hat{c}^{\dagger}_{i\sigma}\hat{c}_{i\sigma}$ and $\hat{n}_{i}=\sum_{\sigma}\hat{n}_{i,\sigma}$ are the electron number operators. The electron hopping amplitude $t_{ij}$ equals $t$ when $i$ and $j$ are the nearest neighbors, and equals $t'$ for next-nearest neighbors. $U$ is the on-site repulsive Coulomb interaction. $V$ is NN electron interaction where $V<0$ and $V>0$ represent electron attraction and repulsion, respectively. We take the lattice geometry to be cylindrical with a lattice spacing of unity. The boundary condition of the cylinders is periodic along the $\hat{y}=(0,1)$ direction and open in the $\hat{x}=(1,0)$ direction. Here, we focus on four-leg cylinders where the width is $L_y=4$ and length up to $L_x=64$, with $L_x$ and $L_y$ are the number of lattice sites along the $\hat{x}$ and $\hat{y}$ directions, respectively. The doped hole concentration is defined as $\delta=N_h/N$, where $N=L_y\times L_x$ is the total number of lattice sites and $N_h$ is the number of doped holes. For the present study, we consider the lightly doped case with hole doping concentration $\delta=12.5\%$. We set $t=1$ as an energy unit, and focus on $U=12$ and $t'=-0.25$ as a representative parameters set. In our calculations, we keep up to $m=16000$ number of states in each DMRG block with a typical truncation error $\epsilon\sim 10^{-6}$.

\begin{figure}[tb]
\centering
    \includegraphics[width=1\linewidth]{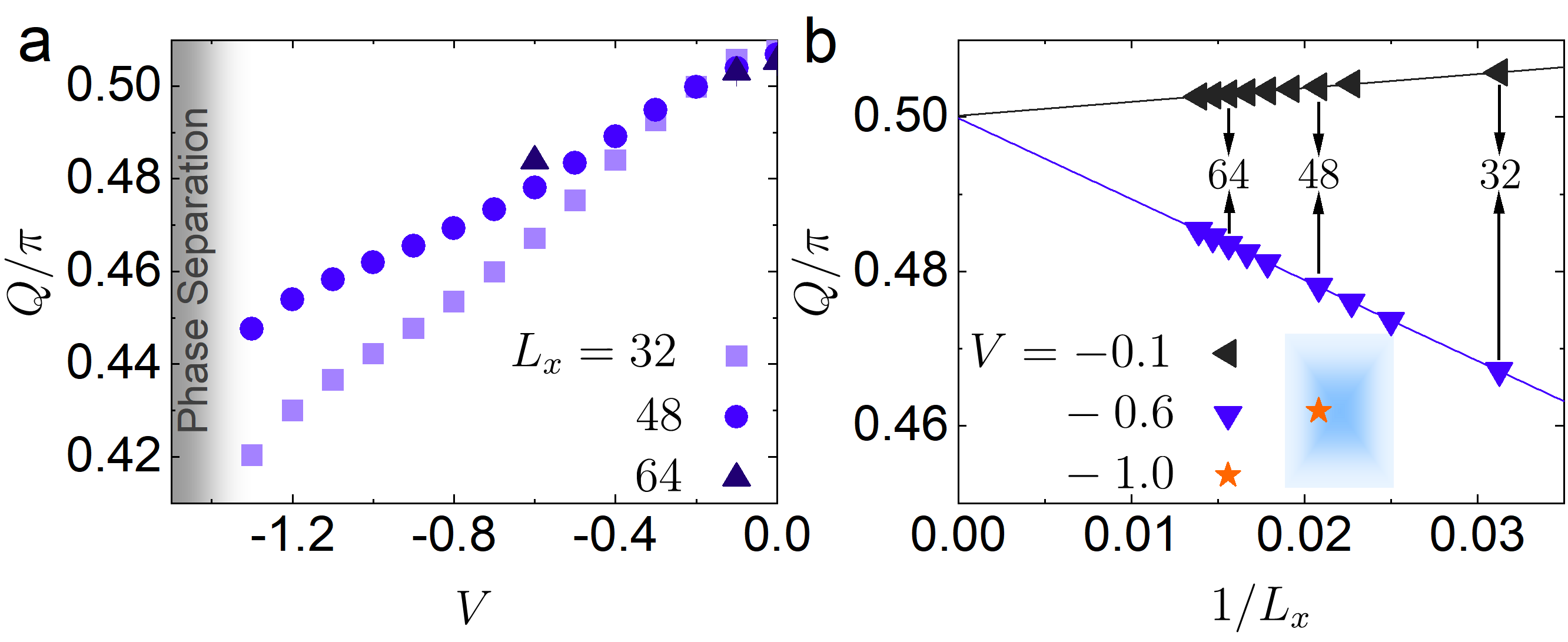}
\caption{(Color online) The CDW ordering wavevector $Q$ as a function of electron attraction $V$ in (\textbf{a}) and the inverse of the cylinder length $L_x$ for different $V$ in (\textbf{b}), respectively. The blue shaded area labels the CDW wavevector $Q=(0.462\pm0.01)\pi$ and charge stripe correlation length $\xi_{\text{co}}= (10.5\pm1.5)\lambda_{\text{c}}$ for La$_{2-x}$Ba$_x$CuO$_4$ with $x=0.125$ \cite{PhysRevB.83.104506}.} \label{Figs:Q-Lx}
\end{figure}

\textit{Charge density wave order --}
To describe the charge density properties of the ground state of the system, we have calculated the charge density profile $n(x,y)=\langle \hat{n}(x,y)\rangle$ and its local rung average $n(x)={\sum_{y=1}^{L_y}} n(x,y)/L_y$. For relatively weak electron attraction $V$, e.g., $V=-0.1$ as shown in the inset of Fig.\ref{Figs:Q-V}, the charge-density distribution $n(x)$ is consistent with the ``half-filled" charge stripe \cite{PRB1997White,PRB1999White,Jiang2018tJ,Jiang2019Hub} of wavelength $\lambda_c=\frac{1}{2\delta}$, i.e., spacings between two adjacent stripes, with half a doped hole per unit cell. This is consistent with previous DMRG studies of the single-band Hubbard model in the absence of electron attraction, i.e., $V=0$ \cite{Jiang2019Hub,Jiang2020prr,Chung2020}. Accordingly, the ordering wavevector $Q=2\pi/\lambda_c$ can be obtained by fitting the charge density oscillation induced by the boundaries of the cylinder\cite{White2002,cdwosc2015prb}
\begin{eqnarray}\label{Eq:FriedelOscillation}
n(x) &\approx& \frac{A\cos(Qx+\phi_{1})}{[L_{\text{eff}} \sin(\pi x/L_{\text{eff}}+\phi_{2})]^{K_c/2}}+n_0.
\end{eqnarray}
Here $A$ is a non-universal amplitude, $\phi_{1}$ and $\phi_{2}$ are the phase shifts, $K_c$ is the Luttinger exponent, and $n_0$ is the mean density. We find that an effective length of $L_{\text{eff}}\sim L_x-2$ best describes our results. As expected, we find that $\lambda_c \sim 4 ~(Q=4\pi\delta \sim \pi/2)$ for the ``half-filled" charge stripe.

When the electron attraction becomes relatively strong, the CDW wavelength $\lambda_{\text{c}}$ starts deviating notably from the``half-filled" charge stripes on finite cylinder (see Fig.\ref{Figs:Q-V} insets). We note that such a deviation for both $\lambda_{\text{c}}$ and $Q$ from their half-filled stripe values becomes smaller with the increase of the length of cylinders, suggesting that this deviation could be a finite-size effect. This is indeed supported by the finite-size scaling of $Q$ as shown in Fig.\ref{Figs:Q-Lx}(b). It is clear that in the long cylinder limit, i.e, $L_x\rightarrow 0$ or $1/L_x\rightarrow 0$, $Q$ for all different $V$ will converge to the same value $Q=4\pi\delta=\pi/2$ where $\lambda_c=1/2\delta=4$ at $\delta=12.5\%$. As a result, the``half-filled" charge stripes may restore in the thermodynamic limit. It is also interesting to note that for cuprates such as La$_{2-x}$Ba$_{x}$CuO$_4$ and La$_{2-x}$Sr$_{x}$CuO$_4$ near $12.5\%$ hole doping, the CDW wavevector $Q$ determined from scattering measurements is always smaller than that expected for the ``half-filled" charge stripe of $4\pi\delta$ \cite{PhysRevB.83.104506,PhysRevB.90.100510,PhysRevB.89.224513,Wen2019}. Considering the fact that experimentally the charge order in cuprates is short-ranged with a correlation length of $\xi_{\text{co}}\approx 11\lambda_{\text{c}}$ in LBCO\cite{PhysRevB.83.104506} and $\xi_{\text{co}}\approx 3\lambda_{\text{c}}$ in LSCO\cite{PhysRevB.90.100510,PhysRevB.89.224513,Wen2019}. Our results shown in Fig.\ref{Figs:Q-Lx}(b) on finite-length cylinders (with $L_x$ comparable to $\xi_{\text{co}}$ in cuprates) are suggestive of significant electron attraction $V$ in these materials, where the estimated values of electron attraction (in the context of single-band Hubbard model) are $V\approx -1.0(2)$ in LBCO and $V\approx -0.40(5)$ in LSCO.

\begin{figure}[tb]
\centering
    \includegraphics[width=1.0\linewidth]{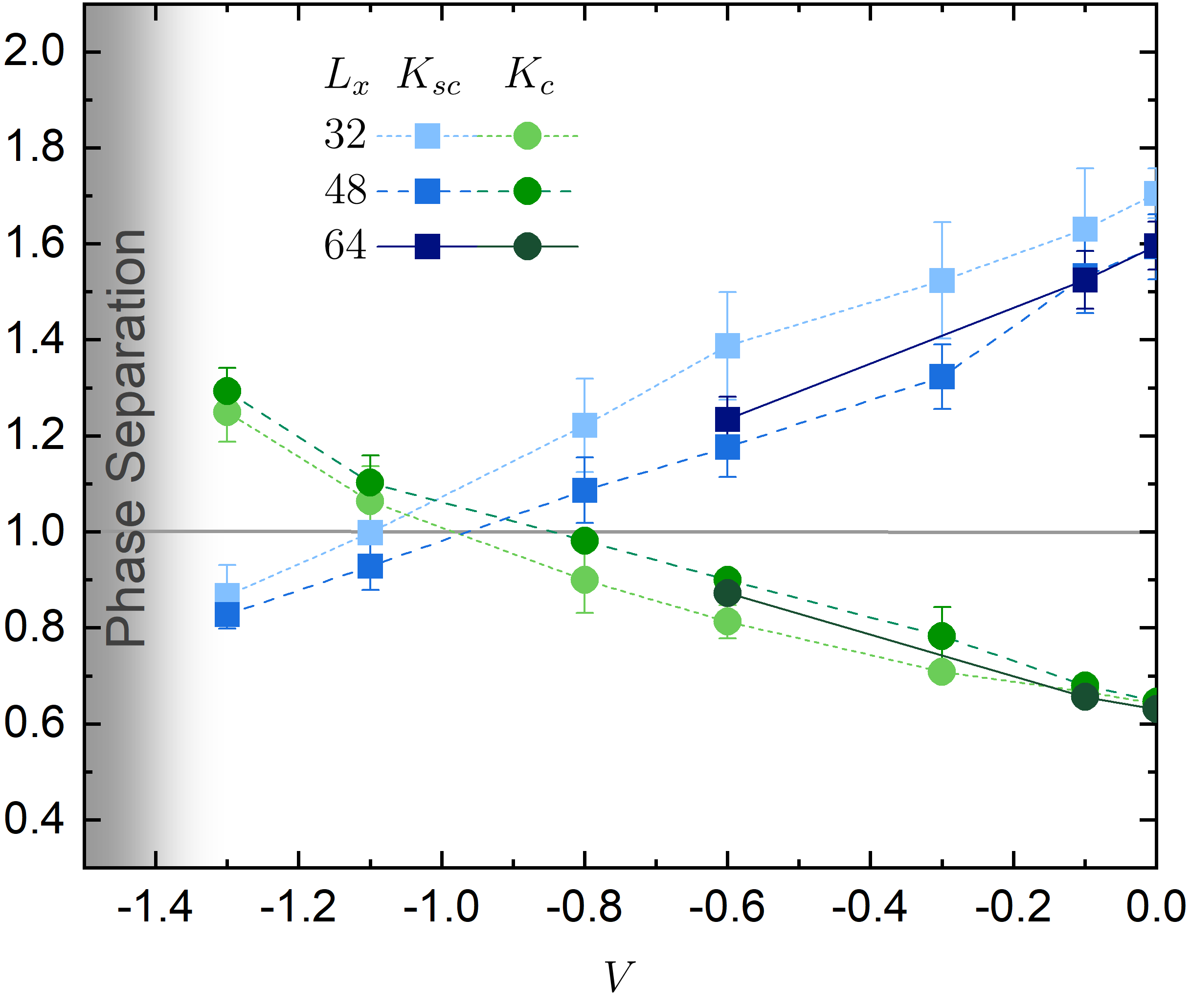}
\caption{(Color online) The extracted Luttinger exponents $K_{sc}$ and $K_c$ as a function of the strength of electron attraction $V$ on different cylinders of length $L_x$. The gray shaded area denotes phase separation.} \label{Figs:Ksc&&Kc-V}
\end{figure}

Importantly, another prominent observation in our study is that while ``half-filled" charge stripes may be restored in the thermodynamic limit, their amplitude and strength is monotonically suppressed by the electron attraction. This is supported by the fact that the CDW exponent $K_c$ defined in Eq. (\ref{Eq:FriedelOscillation}) increases notably with the increase of electron attraction $|V|$ as shown in Fig.\ref{Figs:Ksc&&Kc-V}. For instance, we find that $K_{c}=0.65(1)$ for $V=-0.1$ while $K_{c}=0.87(1)$ for $V=-0.6$ on the four-leg cylinder of length $L_x=64$, with an apparent insensitivity to the length of the ladders used in this study. It is worth mentioning that when $V\lesssim -1.0$ the CDW order is sufficiently suppressed to be secondary, where the SC correlation becomes dominant, as shown in Fig. \ref{Figs:Ksc&&Kc-V}. 

\begin{figure}[htbp!]
\centering
    \includegraphics[width=1\linewidth]{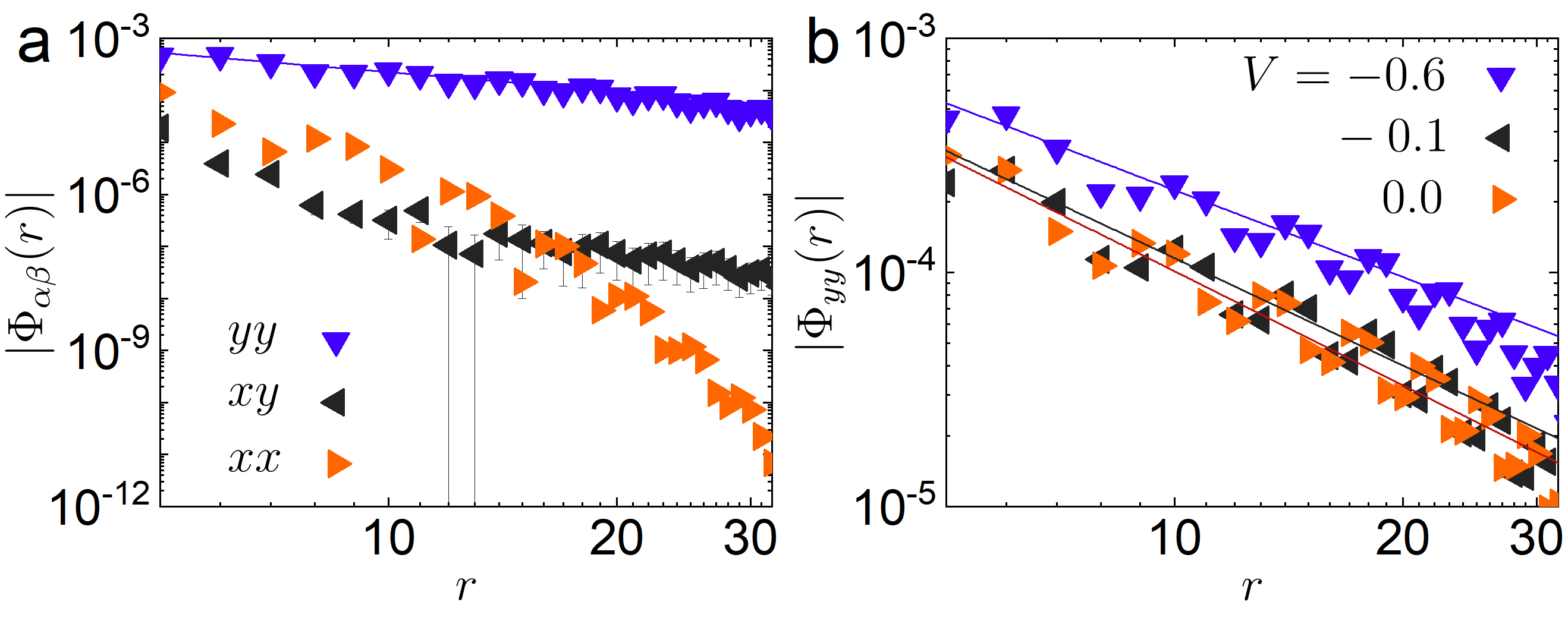}
\caption{(Color online) The superconducting pair-field correlation function for $V=-0.6$ in (\textbf{a}) and charge density profile for $V=-0.6$, $-0.1$ and $0.0$ in (\textbf{b}). The cylinder length is $L_x=64$ and the solid lines are power-law fitting using $\Phi_{yy}(r)\sim r^{-K_{sc}}$.} \label{Figs:sc&&cdw_64x4}
\end{figure}

\textit{Superconducting correlation --}
To describe the superconducting properties of the ground state of the system, we have calculated the equal-time spin-singlet SC correlation function%
\begin{eqnarray}\label{Eq:SC_cor}
\Phi_{\alpha\beta}(r,y)=\langle\Delta^{\dagger}_{\alpha}(x_0,y_0)\Delta_{\beta}(x_0+r,y_0+y)\rangle.
\end{eqnarray}
Here, $\Delta^{\dagger}_{\alpha}(x,y)=\frac{1}{\sqrt{2}}[\hat{c}^{\dagger}_{(x,y),\uparrow}\hat{c}^{\dagger}_{(x,y)+\alpha,\downarrow}-\hat{c}^{\dagger}_{(x,y),\downarrow}\hat{c}^{\dagger}_{(x,y)+\alpha,\uparrow}]$ is spin-singlet pair creation operator on the bond $\alpha=\hat{x}$ or $\hat{y}$. ($x_0,y_0$) is the reference bond with $x_0\sim L_x/4$, $r$ is the distance between two bonds in the $\hat{x}$ direction and $y$ is the displacement between two bonds in the $\hat{y}$ direction. We have calculated different components of the SC correlation, including $\Phi_{xx}$, $\Phi_{xy}$ and $\Phi_{yy}$ and find that the pairing symmetry is consistent with the plaquette $d$-wave symmetry \cite{Dodaro2017,Chung2020}. This is characterized by $|\Phi_{yy}(r,0)|\gg |\Phi_{xy}(r,0)|\gg |\Phi_{xx}(r,0)|$ and $\Phi_{yy}(r,0)=-\Phi_{yy}(r,1)$.

Similar to the CDW correlation, at long distances, $\Phi_{yy}(r)$ is characterized by a power law as shown in Fig.\ref{Figs:sc&&cdw_64x4}(a) with the appropriate Luttinger exponent $K_{sc}$ defined by%
\begin{eqnarray}\label{Eq:SC}
\Phi_{yy}(r)\sim r^{-K_{sc}}.
\end{eqnarray}
As mentioned, previous studies\cite{Jiang2019Hub,Jiang2020prr,Chung2020} on 4-leg square cylinders have shown that without electron attraction, that is, $V=0$, the CDW correlations dominate over the SC correlations as $K_c<K_{sc}$. This suggests that the CDW order may be realized in the 2D limit, although the SC correlations are substantial. It is hence highly nontrivial to find a way to enhance the SC correlations while suppressing the CDW order.

In the previous section, we have shown that CDW correlations can be notably suppressed by NN electron attraction $V$. Accordingly, we would expect that the SC correlations can be enhanced by the NN electron attraction $V$ since the CDW and SC orders are mutually competing \cite{Fradkin2015}. Our numerical results are indeed consistent with this expectation. As shown in Fig.\ref{Figs:Ksc&&Kc-V}, SC correlations become dominant over CDW correlations when $V\lesssim-1.0$, where$K_{sc}< K_c$. While a slow decay of SC correlations with $K_{sc}<2$ implies a SC susceptibility that diverges as $\xi_{sc}\sim T^{-(2-K_{sc})}$ as the temperature $T\rightarrow 0$, a much smaller $K_{sc}$, i.e., $K_{sc}\sim 1$, would lead to a much more divergent SC susceptibility. This suggests that the SC order, instead of the CDW order, could be realized in the 2D limit. As far as we know, to date, this is the first time that the dominant SC correlation has been observed via DMRG in the uniform Hubbard model on the square lattice of width $L_y>2$. It is worth mentioning that while adding a near-neighbor attraction flips the dominant order, the ground state of the system is still consistent with that of a LE liquid phase where $K_{sc}K_{c}\sim 1$. 

\begin{figure}[tb]
\centering
    \includegraphics[width=1\linewidth]{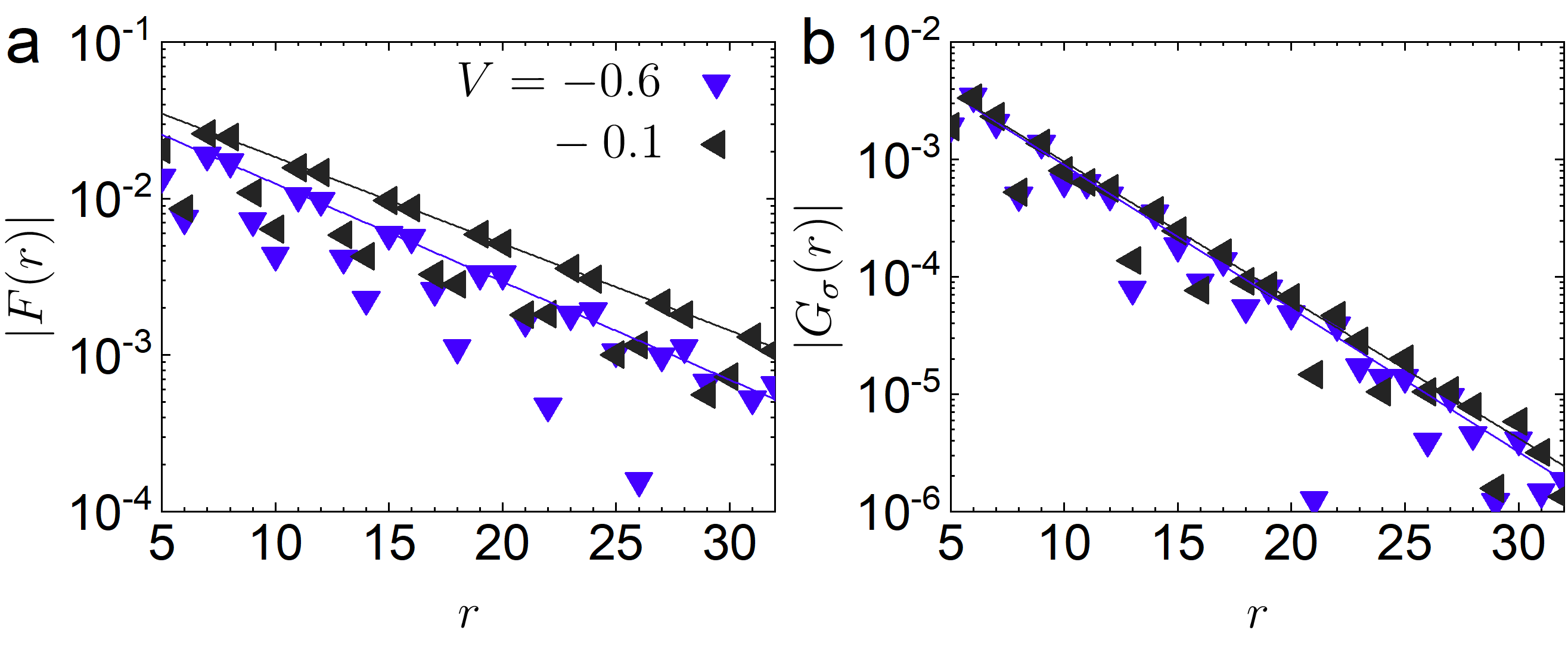}
\caption{(Color online) (\textbf{a}) The spin-spin and (\textbf{b}) single-particle correlation functions with electron attraction $V=-0.6$ and $V=-0.1$. The length of cylinder is $L_x=64$. Solid lines denote exponential fit $F(r)\sim e^{-r/\xi_s}$ and $G_\sigma(r)\sim e^{-r/\xi_G}$, where $\xi_s$ and $\xi_G$ are corresponding correlation lengths.} \label{Figs:spin&&char_64x4}
\end{figure}

\textit{Spin-spin and single-particle correlations --} %
To describe the magnetic properties of the ground state, we calculate the spin-spin correlation function defined as%
\begin{eqnarray}\label{Eq:SpinCor}
F(r)=\langle \vec{S}_{x_0,y_0}\cdot \vec{S}_{x_0+r,y_0}\rangle.
\end{eqnarray}
Here $\vec{S}_{x,y}$ is the spin operator on site $i=(x,y)$ and $i_0=(x_0,y_0)$ is the reference site with $x_0\sim L_x/4$. Fig.\ref{Figs:spin&&char_64x4}(\textbf{a}) shows $F(r)$ for four-leg cylinder of length $L_x=64$. It is clear that $F(r)$ decays exponentially as $F(r)\sim e^{-r/\xi_s}$ at long-distances, with a finite correlation length $\xi_s$. This is consistent with a finite excitation gap in the spin sector. Moreover, we find that $\xi_s$ decreases with increasing $|V$, e.g., $\xi_{s}=7.8(1)$ for $V=-0.1$ and $\xi_{s}=7.0(2)$ for $V=-0.6$. This suppression of short-range antiferromagnetic correlations may thus help to destabilize CDW order and promote SC order. Consistent with previous studies\cite{Jiang2019Hub,Jiang2020prr,Chung2020}, $F(r)$ displays spatial modulations with wavelengths twice that of the charge.

We have also calculated the single-particle Green function, defined as%
\begin{eqnarray}\label{Eq:CC}
G_\sigma(r)=\langle c^{\dagger}_{x_0,y_0,\sigma} c_{x_0+r,y_0,\sigma}\rangle.
\end{eqnarray}
Examples of $G_\sigma(r)$ are shown in Fig.\ref{Figs:spin&&char_64x4}(\textbf{b}). The long distance behavior of $G_\sigma(r)$ is consistent with exponential decay $G_\sigma(r)\sim e^{-r/\xi_G}$. Similar to $\xi_s$, we find that $\xi_G$ also decreases with the increase of $|V|$. For instance, the extracted correlation lengths are $\xi_{G}=3.7(1)$ for $V=-0.1$ and $\xi_{G}=3.5(1)$ for $V=-0.6$, respectively. These are consistent with that of the LE phase.

\textit{Summary and discussion --}
In this paper, we have studied the ground state properties of the lightly doped extended Hubbard model on the four-leg square cylinders in the presence of near-neighbor electron attraction. Taken together, our results show that the ground state of the system is consistent with a LE liquid \cite{Luther1974} where both CDW and SC correlations decay as a power-law and $K_{sc}K_c\sim 1$. However, previous studies\cite{Jiang2019Hub,Jiang2020prr,Chung2020} show that SC correlations are secondary when $V=0$ compared with CDW correlations since $K_c<1<K_{sc}$. Interestingly, we find that the near-neighbor electron attraction $V$ can significantly enhance SC correlations while simultaneously suppress CDW correlations. As a result, when $V\lesssim -1.0$, SC correlations become dominant while CDW correlations become secondary as $K_{sc}<1<K_c$. To the best of our knowledge, to date, this is the first numerical realization of dominant superconductivity in doping the uniform Hubbard model on the square lattice of width wider than a 2-leg ladder. While in this paper we have focused on the effect of electron attraction $V$ on the LE liquid phase, it will also be interesting to study its effect on doping a qualitatively distinct phase, such as the insulating ``filled" stripe phases\cite{Zaanen1989prb} with $t'=0$ on the Hubbard model, and see whether superconductivity can likewise be obtained. Answering these questions may lead to better understanding of the mechanism of high-temperature superconductivity.

We note that the critical $V_c$, where superconductivity starts to dominate in our simulation, is consistent with the recently identified attractive interaction in 1D cuprates Ba$_{2-x}$Sr$_x$CuO$_{3+\delta}$.\cite{chen2021anomalously} Considering the chemical similarity, it is reasonable to believe that the effective near-neighbor attraction in the CuO$_2$ plane is comparable to $V\sim -t$. Therefore, our finding suggests the importance of additional interactions beyond the Hubbard model to stabilize superconductivity over CDW. Future high-resolution experiments, such as photoemission and x-ray scattering, and their comparisons with numerical simulations may quantify this effective interaction $V$ in high-$T_c$ cuprates, in a similar way as for the 1D cuprate chains.

Another approach to estimating this effective $V$ in realistic materials is a microscopic analysis of cuprates' crystal and electronic structure. Site phonons coupled with electronic density are possible candidates to mediate such an attractive interaction, which has been discussed quantitatively\,\cite{wang2021phonon}. However, a combined impact of other phonons and other bosonic excitations may contribute to this effective interaction and ultimately result in the strong $d$-wave superconductivity in cuprates.

We would like to thank Steven Kivelson for insightful discussions. This work was supported by the Department of Energy, Office of Science, Basic Energy Sciences, Materials Sciences and Engineering Division, under Contract No. DE-AC02-76SF00515. Parts of the computing for this project were performed on the National Energy Research Scientific Computing Center (NERSC), a US Department of Energy Office of Science User Facility operated under Contract No. DE-AC02-05CH11231. Parts of the computing for this project were performed on the Sherlock cluster. Y.W. acknowledges support from the National Science Foundation (NSF) award DMR-2132338.


\begin{thebibliography}{10}

\bibitem{bednorz1986possible}
J~George Bednorz and K~Alex M{\"u}ller.
\newblock \textit{Possible High-T$_C$ Superconductivity in the Ba-La-Cu-O
  System}.
\newblock {\em Z. Phys. B Condens. Matter}, 64(2):189, 1986.

\bibitem{scalapino1986d}
DJ~Scalapino, E~Loh~Jr, and JE~Hirsch.
\newblock \textit{d-Wave Pairing near a Spin-Density-Wave Instability}.
\newblock {\em Phys. Rev. B}, 34(11):8190, 1986.

\bibitem{gros1987superconducting}
C~Gros, R~Joynt, and TM~Rice.
\newblock \textit{Superconducting Instability in the Large-U Limit of the
  Two-Dimensional Hubbard Model}.
\newblock {\em Z. Phys. B Condens. Matter}, 68(4):425, 1987.

\bibitem{kotliar1988superexchange}
Gabriel Kotliar and Jialin Liu.
\newblock \textit{Superexchange Mechanism and d-Wave Superconductivity}.
\newblock {\em Phys. Rev. B}, 38(7):5142, 1988.

\bibitem{tsuei2000pairing}
CC~Tsuei and JR~Kirtley.
\newblock \textit{Pairing Symmetry in Cuprate Superconductors}.
\newblock {\em Rev. Mod. Phys.}, 72(4):969, 2000.

\bibitem{zhang1988effective}
FC~Zhang and TM~Rice.
\newblock Effective hamiltonian for the superconducting cu oxides.
\newblock {\em Phys. Rev. B}, 37(7):3759, 1988.

\bibitem{natphyhubbard2013}
Editorial.
\newblock The hubbard model at half a century.
\newblock {\em Nat. Phys.}, 9:523, 2013.

\bibitem{Keimer2015nat}
B.~Keimer, S.~A. Kivelson, M.~R. Norman, S.~Uchida, and J.~Zaanen.
\newblock From quantum matter to high-temperature superconductivity in copper
  oxides.
\newblock {\em Nature}, 518:179–186, 2015.

\bibitem{Srinivas2022Hubbard}
Daniel~P. Arovas, Erez Berg, Steven~A. Kivelson, and Srinivas Raghu.
\newblock The hubbard model.
\newblock {\em Annual Review of Condensed Matter Physics}, 13(1):239--274,
  2022.

\bibitem{Gull2021Hubbard}
Mingpu Qin, Thomas Schäfer, Sabine Andergassen, Philippe Corboz, and Emanuel
  Gull.
\newblock The hubbard model: A computational perspective.
\newblock {\em Annual Review of Condensed Matter Physics}, 13(1):275--302,
  2022.

\bibitem{anderson1987resonating}
Philip~W Anderson.
\newblock \textit{The Resonating Valence Bond State in La$_2$CuO$_4$ and
  superconductivity}.
\newblock {\em Science}, 235(4793):1196, 1987.

\bibitem{dagotto1994correlated}
Elbio Dagotto.
\newblock \textit{Correlated Electrons in High-Temperature Superconductors}.
\newblock {\em Rev. Mod. Phys.}, 66(3):763, 1994.

\bibitem{zheng2017stripe}
Bo-Xiao Zheng, Chia-Min Chung, Philippe Corboz, Georg Ehlers, Ming-Pu Qin,
  Reinhard~M Noack, Hao Shi, Steven~R White, Shiwei Zhang, and Garnet Kin-Lic
  Chan.
\newblock {Stripe Order in the Underdoped Region of the Two-Dimensional Hubbard
  Model}.
\newblock {\em Science}, 358(6367):1155, 2017.

\bibitem{huang2017numerical}
Edwin~W Huang, Christian~B Mendl, Shenxiu Liu, Steve Johnston, Hong-Chen Jiang,
  Brian Moritz, and Thomas~P Devereaux.
\newblock {Numerical Evidence of Fluctuating Stripes in the Normal State of
  High-Tc Cuprate Superconductors}.
\newblock {\em Science}, 358(6367):1161, 2017.

\bibitem{huang2017science}
Edwin~W. Huang, Christian~B. Mendl, Shenxiu Liu, Steve Johnston, Hong-Chen
  Jiang, Brian Moritz, and Thomas~P. Devereaux.
\newblock Numerical evidence of fluctuating stripes in the normal state of
  high-<i>t</i><sub>c</sub> cuprate superconductors.
\newblock {\em Science}, 358(6367):1161--1164, 2017.

\bibitem{huang2018stripe}
Edwin~W Huang, Christian~B Mendl, Hong-Chen Jiang, Brian Moritz, and Thomas~P
  Devereaux.
\newblock {Stripe Order from the Perspective of the Hubbard Model}.
\newblock {\em npj Quantum Mater.}, 3(1):22, 2018.

\bibitem{ponsioen2019period}
Boris Ponsioen, Sangwoo~S Chung, and Philippe Corboz.
\newblock {Period 4 Stripe in the Extended Two-Dimensional Hubbard Model}.
\newblock {\em Phys. Rev. B}, 100(19):195141, 2019.

\bibitem{kokalj2017bad}
Jure Kokalj.
\newblock {Bad-Metallic Behavior of Doped Mott Insulators}.
\newblock {\em Phys. Rev. B}, 95(4):041110, 2017.

\bibitem{huang2019strange}
Edwin~W Huang, Ryan Sheppard, Brian Moritz, and Thomas~P Devereaux.
\newblock {Strange Metallicity in the Doped Hubbard Model}.
\newblock {\em Science}, 366(6468):987, 2019.

\bibitem{cha2020slope}
Peter Cha, Aavishkar~A Patel, Emanuel Gull, and Eun-Ah Kim.
\newblock {Slope Invariant T-Linear Resistivity from Local Self-Energy}.
\newblock {\em Phys. Rev. Research}, 2(3):033434, 2020.

\bibitem{Mingpu2020prx}
Mingpu Qin, Chia-Min Chung, Hao Shi, Ettore Vitali, Claudius Hubig, Ulrich
  Schollw\"ock, Steven~R. White, and Shiwei Zhang.
\newblock Absence of superconductivity in the pure two-dimensional hubbard
  model.
\newblock {\em Phys. Rev. X}, 10:031016, Jul 2020.

\bibitem{Jiang2019Hub}
Hong-Chen Jiang and Thomas~P. Devereaux.
\newblock Superconductivity in the doped hubbard model and its interplay with
  next-nearest hopping t'.
\newblock {\em Science}, 365(6460):1424--1428, 2019.

\bibitem{Jiang2020prr}
Yi-Fan Jiang, Jan Zaanen, Thomas~P. Devereaux, and Hong-Chen Jiang.
\newblock Ground state phase diagram of the doped hubbard model on the four-leg
  cylinder.
\newblock {\em Phys. Rev. Research}, 2:033073, Jul 2020.

\bibitem{Chung2020}
Chia-Min Chung, Mingpu Qin, Shiwei Zhang, Ulrich Schollw\"ock, and Steven~R.
  White.
\newblock Plaquette versus ordinary $d$-wave pairing in the
  ${t}^{\ensuremath{'}}$-hubbard model on a width-4 cylinder.
\newblock {\em Phys. Rev. B}, 102:041106, Jul 2020.

\bibitem{Dodaro2017}
John~F. Dodaro, Hong-Chen Jiang, and Steven~A. Kivelson.
\newblock {Intertwined order in a frustrated four-leg $t$-$J$ cylinder}.
\newblock {\em Phys. Rev. B}, 95:155116, Apr 2017.

\bibitem{Jiang2018tJ}
Hong-Chen Jiang, Zheng-Yu Weng, and Steven~A. Kivelson.
\newblock {Superconductivity in the doped $t$-$J$ model: Results for four-leg
  cylinders}.
\newblock {\em Phys. Rev. B}, 98:140505, Oct 2018.

\bibitem{jiang2021ground}
Shengtao Jiang, Douglas~J Scalapino, and Steven~R White.
\newblock \textit{Ground State Phase Diagram of the $ t $-$ t'$-$ J $ Model}.
\newblock {\em Proc. Natl. Acad. Sci. U.S.A.}, 118:e2109978118, 2021.

\bibitem{Jiang2021prl}
Hong-Chen Jiang and Steven~A. Kivelson.
\newblock High temperature superconductivity in a lightly doped quantum spin
  liquid.
\newblock {\em Phys. Rev. Lett.}, 127:097002, Aug 2021.

\bibitem{Gong2021}
Shoushu Gong, W.~Zhu, and D.~N. Sheng.
\newblock Robust $d$-wave superconductivity in the square-lattice
  $t\text{\ensuremath{-}}j$ model.
\newblock {\em Phys. Rev. Lett.}, 127:097003, Aug 2021.

\bibitem{Jiang2020prb}
Hong-Chen Jiang, Shuai Chen, and Zheng-Yu Weng.
\newblock Critical role of the sign structure in the doped mott insulator:
  Luther-emery versus fermi-liquid-like state in quasi-one-dimensional ladders.
\newblock {\em Phys. Rev. B}, 102:104512, Sep 2020.

\bibitem{Jiang2022pnas}
Hong-Chen Jiang and Steven~A. Kivelson.
\newblock Stripe order enhanced superconductivity in the hubbard model.
\newblock {\em Proceedings of the National Academy of Sciences},
  119(1):e2109406119, 2022.

\bibitem{chen2021anomalously}
Zhuoyu Chen, Yao Wang, Slavko~N. Rebec, Tao Jia, Makoto Hashimoto, Donghui Lu,
  Brian Moritz, Robert~G. Moore, Thomas~P. Devereaux, and Zhi-Xun Shen.
\newblock Anomalously strong near-neighbor attraction in doped 1d cuprate
  chains.
\newblock {\em Science}, 373(6560):1235--1239, 2021.

\bibitem{wang2021phonon}
Yao Wang, Zhuoyu Chen, Tao Shi, Brian Moritz, Zhi-Xun Shen, and Thomas~P
  Devereaux.
\newblock Phonon-mediated long-range attractive interaction in one-dimensional
  cuprates.
\newblock {\em Phys. Rev. Lett.}, 127(19):197003, 2021.

\bibitem{jiang2022enhancing}
Mi~Jiang.
\newblock Enhancing d-wave superconductivity with nearest-neighbor attraction
  in the extended hubbard model.
\newblock {\em Physical Review B}, 105(2):024510, 2022.

\bibitem{huang2021superconductivity}
Zhong-Bing Huang, Shi-Chao Fang, and Hai-Qing Lin.
\newblock Superconductivity, nematicity, and charge density wave in high-tc
  cuprates: A common thread.
\newblock {\em arXiv:2109.05519}, 2021.

\bibitem{Qu2021arxiv}
Dai-Wei {Qu}, Bin-Bin {Chen}, Hong-Chen {Jiang}, Yao {Wang}, and Wei {Li}.
\newblock {Spin-Triplet Pairing Induced by Near-Neighbor Attraction in the
  Cuprate Chain}.
\newblock {\em arXiv e-prints}, page arXiv:2110.00564, October 2021.

\bibitem{Luther1974}
A.~Luther and V.~J. Emery.
\newblock Backward scattering in the one-dimensional electron gas.
\newblock {\em Phys. Rev. Lett.}, 33:589--592, Sep 1974.

\bibitem{White1992}
Steven~R. White.
\newblock Density matrix formulation for quantum renormalization groups.
\newblock {\em Phys. Rev. Lett.}, 69:2863--2866, Nov 1992.

\bibitem{PhysRevB.83.104506}
M.~H\"ucker, M.~v.~Zimmermann, G.~D. Gu, Z.~J. Xu, J.~S. Wen, Guangyong Xu,
  H.~J. Kang, A.~Zheludev, and J.~M. Tranquada.
\newblock Stripe order in superconducting {La$_{2-x}$Ba$_{x}$CuO$_{4}$ ($0.095$
  $\ensuremath{\leq}$ $x$ $\ensuremath{\leq$} $0.155$)}.
\newblock {\em Phys. Rev. B}, 83:104506, Mar 2011.

\bibitem{PRB1997White}
Steven~R. White and D.~J. Scalapino.
\newblock Ground states of the doped four-leg t-j ladder.
\newblock {\em Phys. Rev. B}, 55:R14701--R14704, Jun 1997.

\bibitem{PRB1999White}
Steven~R. White and D.~J. Scalapino.
\newblock Competition between stripes and pairing in a
  ${t\ensuremath{-}t}^{\ensuremath{'}}\ensuremath{-}j$ model.
\newblock {\em Phys. Rev. B}, 60:R753--R756, Jul 1999.

\bibitem{White2002}
Steven~R. White, Ian Affleck, and Douglas~J. Scalapino.
\newblock Friedel oscillations and charge density waves in chains and ladders.
\newblock {\em Phys. Rev. B}, 65:165122, Apr 2002.

\bibitem{cdwosc2015prb}
Michele Dolfi, Bela Bauer, Sebastian Keller, and Matthias Troyer.
\newblock Pair correlations in doped hubbard ladders.
\newblock {\em Phys. Rev. B}, 92:195139, Nov 2015.

\bibitem{PhysRevB.90.100510}
V.~Thampy, M.~P.~M. Dean, N.~B. Christensen, L.~Steinke, Z.~Islam, M.~Oda,
  M.~Ido, N.~Momono, S.~B. Wilkins, and J.~P. Hill.
\newblock Rotated stripe order and its competition with superconductivity in
  {${\mathrm{La}}_{1.88}{\mathrm{Sr}}_{0.12}{\mathrm{CuO}}_{4}$}.
\newblock {\em Phys. Rev. B}, 90:100510(R), Sep 2014.

\bibitem{PhysRevB.89.224513}
T.~P. Croft, C.~Lester, M.~S. Senn, A.~Bombardi, and S.~M. Hayden.
\newblock Charge density wave fluctuations in
  {${\text{La}}_{2\ensuremath{-}x}$${\text{Sr}}_{x}$${\text{CuO}}_{4}$} and
  their competition with superconductivity.
\newblock {\em Phys. Rev. B}, 89:224513, Jun 2014.

\bibitem{Wen2019}
J.-J. Wen, H.~Huang, S.-J. Lee, H.~Jang, J.~Knight, Y.~S. Lee, M.~Fujita, K.~M.
  Suzuki, S.~Asano, S.~A. Kivelson, C.-C. Kao, and J.-S. Lee.
\newblock Observation of two types of charge-density-wave orders in
  superconducting {La$_{2-x}$Sr$_x$CuO$_4$}.
\newblock {\em Nat. Commun.}, 10(1):3269, 2019.

\bibitem{Fradkin2015}
Eduardo {Fradkin}, Steven~A. {Kivelson}, and John~M. {Tranquada}.
\newblock {Colloquium: Theory of intertwined orders in high temperature
  superconductors}.
\newblock {\em Reviews of Modern Physics}, 87(2):457--482, April 2015.

\bibitem{Zaanen1989prb}
Jan Zaanen and Olle Gunnarsson.
\newblock Charged magnetic domain lines and the magnetism of high-${T}_{c}$
  oxides.
\newblock {\em Phys. Rev. B}, 40:7391--7394, Oct 1989.

\end{thebibliography}

\end{document}